\documentclass[conference]{IEEEtran}
\IEEEoverridecommandlockouts

\usepackage{graphicx}
\usepackage{textcomp}
\usepackage{xcolor}
\usepackage[backend=biber,style=ieee]{biblatex}
\usepackage{amsmath,amsfonts,amssymb,amsthm}
\usepackage{bm}
\usepackage{mathrsfs}
\usepackage{extarrows}
\usepackage{mathtools}
\usepackage{dsfont}
\usepackage{xparse}
\usepackage{cleveref}
\usepackage{cancel}

\crefformat{equation}{(#2#1#3)}
\crefrangeformat{equation}{(#3#1#4) to~(#5#2#6)}

\NewDocumentCommand{\prox}{g}{%
    \IfNoValueTF{#1}%
    {
        \operatorname{prox}
    }%
    {%
        \operatorname{prox}\left( #1 \right)
    }%
}
\NewDocumentCommand{\blkdiag}{g}{%
    \IfNoValueTF{#1}%
    {
        \operatorname{blkdiag}
    }%
    {%
        \operatorname{blkdiag}\left( #1 \right)
    }%
}
\NewDocumentCommand{\sinc}{g}{%
    \IfNoValueTF{#1}%
    {
        \operatorname{sinc}
    }%
    {%
        \operatorname{sinc}\left( #1 \right)
    }%
}
\RenewDocumentCommand{\Re}{g}{%
    \IfNoValueTF{#1}%
    {
        \operatorname{Re}
    }%
    {%
        \operatorname{Re}\left( #1 \right)
    }%
}
\RenewDocumentCommand{\Im}{g}{%
    \IfNoValueTF{#1}%
    {
        \operatorname{Im}
    }%
    {%
        \operatorname{Im}\left( #1 \right)
    }%
}

\NewDocumentCommand{\Beta}{g}{%
    \IfNoValueTF{#1}%
    {
        \operatorname{Beta}
    }%
    {%
        \operatorname{Beta}\left( #1 \right)
    }%
}

\NewDocumentCommand{\stack}{g}{%
    \IfNoValueTF{#1}%
    {
        \operatorname{stack}
    }%
    {%
        \operatorname{stack}\left( #1 \right)
    }%
}

\NewDocumentCommand{\extr}{g}{%
    \IfNoValueTF{#1}%
    {
        \operatorname{extr}
    }%
    {%
        \operatorname{extr}\left( #1 \right)
    }%
}

\NewDocumentCommand{\rank}{g}{%
    \IfNoValueTF{#1}%
    {
        \operatorname{rank}
    }%
    {%
        \operatorname{rank}\left( #1 \right)
    }%
}
\NewDocumentCommand{\trace}{g}{%
    \IfNoValueTF{#1}%
    {
        \operatorname{trace}
    }%
    {%
        \operatorname{trace}\left( #1 \right)
    }%
}

\NewDocumentCommand{\sign}{g}{%
    \IfNoValueTF{#1}%
    {
        \operatorname{sign}
    }%
    {%
        \operatorname{sign}\left( #1 \right)
    }%
}

\newcommand{\diag}[1]{\operatorname{diag}\left( #1 \right)}

\RenewDocumentCommand{\vec}{g}{%
    \IfNoValueTF{#1}
    { \operatorname{vec} }%
    { \operatorname{vec}\left( #1 \right) }%
}
\NewDocumentCommand{\supp}{g}{%
    \IfNoValueTF{#1}
    { \operatorname{supp} }%
    { \operatorname{supp}\left( #1 \right) }%
}
\NewDocumentCommand{\MSE}{g}{%
    \IfNoValueTF{#1}
    { \operatorname{MSE} }%
    { \operatorname{MSE}\left[ #1 \right] }%
}

\RenewDocumentCommand{\i}{gg}{%
    \IfNoValueTF{#2}%
    {
        \IfNoValueTF{#1}
        { i }%
        { i({#1} \vert \MR) }
    }%
    {%
        i({#2} \vert {#1})
    }%
}

\newcommand{\boldA}{\mathbf{A}}

\newcommand{\boldD}{\mathbf{D}}

\newcommand{\boldF}{\mathbf{F}}

\newcommand{\boldH}{\mathbf{H}}
\newcommand{\boldI}{\mathbf{I}}
\newcommand{\boldJ}{\mathbf{J}}

\newcommand{\boldN}{\mathbf{N}}

\newcommand{\boldP}{\mathbf{P}}

\newcommand{\boldR}{\mathbf{R}}

\newcommand{\boldV}{\mathbf{V}}

\newcommand{\boldY}{\mathbf{Y}}

\newcommand{\bolda}{\mathbf{a}}

\newcommand{\bolde}{\mathbf{e}}
\newcommand{\boldf}{\mathbf{f}}

\newcommand{\boldn}{\mathbf{n}}

\newcommand{\boldp}{\mathbf{p}}

\newcommand{\boldx}{\mathbf{x}}
\newcommand{\boldy}{\mathbf{y}}

\newcommand{\rmc}{\mathrm{c}}
\newcommand{\rmd}{\mathrm{d}}

\newcommand{\rmg}{\mathrm{g}}

\newcommand{\rmm}{\mathrm{m}}
\newcommand{\rmn}{\mathrm{n}}

\newcommand{\rmr}{\mathrm{r}}
\newcommand{\rms}{\mathrm{s}}
\newcommand{\rmt}{\mathrm{t}}

\newcommand{\rmy}{\mathrm{y}}

\newcommand{\calB}{\mathcal{B}}
\newcommand{\calC}{\mathcal{C}}

\newcommand{\calG}{\mathcal{G}}
\newcommand{\calH}{\mathcal{H}}

\newcommand{\calL}{\mathcal{L}}

\newcommand{\calN}{\mathcal{N}}

\newcommand{\calU}{\mathcal{U}}

\newcommand{\bbC}{\mathbb{C}}

\newcommand{\bbE}{\mathbb{E}}

\newcommand{\boldzero}{\mathbf{0}}

\newcommand{\boldalpha}{\bm{\alpha}}

\newcommand{\boldeta}{\bm{\eta}}

\newcommand{\NS}{N_{\rms}}

\renewcommand{\NG}{N_{\rmg}}
\newcommand{\NT}{N_{\rmt}}

\newcommand{\nT}{n_{\rmt}}

\newcommand{\nS}{n_{\rms}}

\newcommand{\nG}{n_{\rmg}}

\newcommand{\mT}{m_{\rmt}}
\newcommand{\mR}{m_{\rmr}}

\newcommand{\MR}{M_{\rmr}}
\newcommand{\MT}{M_{\rmt}}

\newcommand{\CRB}{{\mathrm{CRB}}}

\usepackage{algorithm}
\usepackage[linesnumbered,lined,boxed,algo2e]{algorithm2e}
\usepackage{booktabs}
\usepackage{cleveref}
\usepackage[per-mode=repeated-symbol]{siunitx}

\crefname{figure}{fig.}{fig.}
\Crefname{figure}{Fig.}{Fig.}
\crefname{table}{tab.}{tab.}
\Crefname{Table}{Tab.}{Tab.}
\def\beamidx{b}
\def\beamIdx{B}
\def\beamset{\calB}

\begin{document}

\title{Moving Target Sensing for ISAC Systems in Clutter Environment
}

\author{\IEEEauthorblockN{Dongqi Luo, Huihui Wu, Hongliang Luo, Bo Lin, and Feifei Gao}
\IEEEauthorblockA{Department of Automation, Tsinghua University, Beijing 100084, China \\
\{dongqiluo,hhwu1994\}@mail.tsinghua.edu.cn, \{luohl23,linb20\}@mails.tsinghua.edu.cn, feifeigao@ieee.org}
}

\maketitle

\begin{abstract}
    In this paper, we consider the moving target sensing problem for integrated sensing and communication (ISAC) systems in clutter environment. Scatterers produce strong clutter, deteriorating the performance of ISAC systems in practice. Given that scatterers are typically stationary and the targets of interest are usually moving, we here focus on sensing the moving targets. Specifically, we adopt a scanning beam to search for moving target candidates. For the received signal in each scan, we employ high-pass filtering in the Doppler domain to suppress the clutter within the echo, thereby identifying candidate moving targets according to the power of filtered signal. Then, we adopt root-MUSIC-based algorithms to estimate the angle, range, and radial velocity of these candidate moving targets. Subsequently, we propose a target detection algorithm to reject false targets. Simulation results validate the effectiveness of these proposed methods.
\end{abstract}

\section{Introduction}
Integrated sensing and communication (ISAC) is a promising technology for the sixth generation (6G) of wireless networks, and has attracted increasing attention in recent years~\cite{Liu20226GBeyond}. By endowing communication devices with the capability to perceive their environment, ISAC not only conserves spectrum resources but also economizes on hardware resources~\cite{Cui2021Ubiquitous,Zhang2021AnOverview}. 

According to the stage of target sensing, existing works on ISAC fall into two categories: transmitting and echo signal processing. In the transmitting stage, the objective is to design appropriate beamformers or symbols that meet both sensing and communication criteria, e.g., the Cramér-Rao bound (CRB), the sum-rate, and the transmitting power~\cite{Liu2020JointTransmit,Zhao2022Joint,Liu2022Cramer,Chen2021Joint}. When tracking high-speed targets, the sensing beam need to be steered towards the forecasted locations~\cite{Yuan2021Bayesian,Liu2022Learning,Zhen2023Integrated}. In the echo signal processing stage, the base station (BS) receives the echo signal, and estimates target parameters using target sensing algorithms~\cite{Gao2023Integrated,Chen2023Multiple,luo2022beam}. 

Prior works predominantly focused on target sensing in free space, while only the reflections of the target of interest are considered in the environment. However, in practical applications, ISAC devices would be populated with stationary scatters in the environments. Although these scatters primarily remain stationary, they nonetheless reflect the sensing signal, and pose significant interference to target sensing. In the literature, there is only a few of works that take the clutter environment into consideration~\cite{Su2021Secure,Liao2023Optimized}. These works employed spatial-domain clutter suppression approaches to combat the clutter environment. However, these clutter mitigation methods require the reflection coefficients and directions of all scatters, which are usually unavailable in practice.

Fortunately, due to the stationarity of the environment, we can sense environmental information in advance without the necessity for frequent updates, which also suggests that we only need to focus on sensing the moving targets. Hence, in this paper, we investigate the problem of sensing moving targets in clutter environment. First, a scanning beam is introduced to search for potential moving targets. Subsequently, for the echo of each scan, we design a Doppler domain high-pass filter to suppress the clutter, enabling the identification of echos that may contain potential moving targets, and leverage root-MUSIC-based algorithms to estimate the kinematic parameters of candidate moving targets. Furthermore, based on the assumption that the clutter falls within a specific subspace, we devise a target detection algorithm to reject falsely identified targets. Numerical results are provided to verify the performance of the proposed methods.

\section{System Model}
\label{sec:system-model}

\begin{figure}[t!]
    \centering
    \includegraphics[width=.9\linewidth]{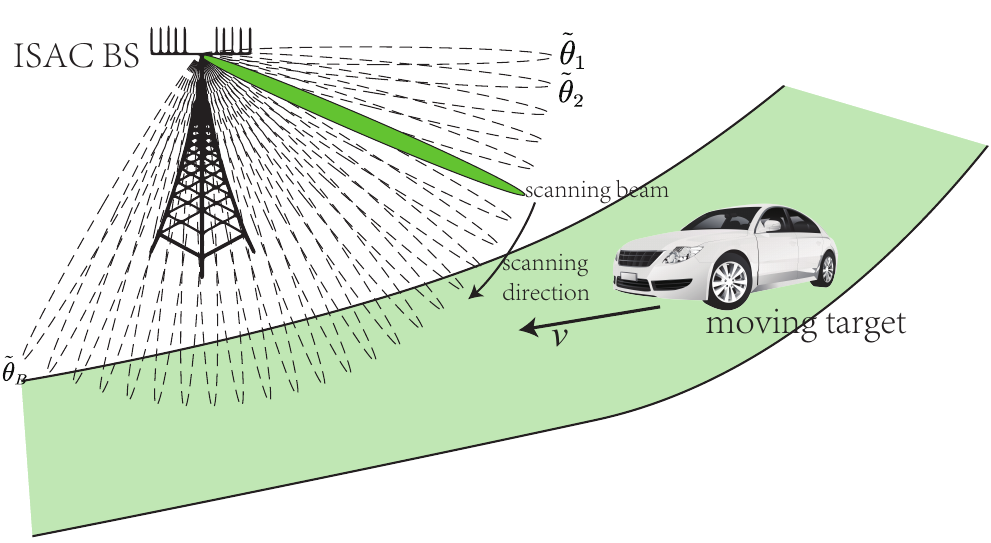}
    \caption{Scanning moving targets with beams.}
    \label{fig:system-model}
\end{figure}

We consider an ISAC system with a transmitting uniform linear array (ULA) of $\MT$ antennas and a receiving ULA of $\MR$ antennas. The BS transmits OFDM signal with $L$ subcarriers for both sensing and communication. The carrier frequency is $f_{\rmc}$, and the wavelength is $\lambda = c / f_{\rmc}$, where $c$ is the light speed. The $l$th subcarrier frequency is $f_{l} = f_{\rmc} + (l - 1) \Delta f$, with $\Delta f$ being the frequency interval. Let $T_{\rms\rmy\rmm}$ be the symbol period, and then $\Delta f = 1 / T_{\rms\rmy\rmm}$. Moreover, we denote $T_{\rmg}$ as the guard interval, and the symbol interval can be expressed as $T = T_{\rms\rmy\rmm} + T_{\rmg}$. As shown in \Cref{fig:system-model}, the ISAC BS adopts a scanning beam to search for candidate moving targets, whose set of beam directions is denoted as $\{ \tilde{\theta}_{1}, \dots, \tilde{\theta}_{\beamIdx} \}$. In the $\beamidx$th scan, the beam is steered towards $\tilde{\theta}_{\beamidx}$, and the corresponding coverage is denoted as $\Theta_{\beamidx}$. We assume that targets are distinct and the beam is narrow, and thus only one target exists within a single scan.

We assume that there are $\NT$ moving targets. The angle, range, and radial speed of the $\nT$th target are denoted as $\theta_{\nT}^{\rmt}$, $r_{\nT}^{\rmt}$ and $v_{\nT}^{\rmt}$, respectively. We define the corresponding range, Doppler and spatial frequency as $\psi_{\rmr, \nT}^{\rmt} = 2r_{\nT}^{\rmt}\Delta f / c$, $\psi_{\rmd, \nT}^{\rmt} = 2 v_{\nT}^{\rmt} T / \lambda$ and $\psi_{\rms, \nT}^{\rmt} = d \sin\theta_{\nT}^{\rmt} / \lambda$. The sensing channel of the $\nT$th target on the $l$th subcarrier at the $p$th symbol is 
\begin{align}
  \boldH_{\nT, l, p}^{\rmt} = \alpha_{\nT}^{\rmt} a_{\rmd, \nT, p}^{\rmt} a_{\rmr, \nT, l}^{\rmt} \boldA_{\rms}(\psi_{\rms, \nT}^{\rmt}),
\end{align}
where $\alpha_{\nT}^{\rmt}$ is the $\nT$th target's reflection coefficient, $a_{\rmd, \nT, p}^{\rmt} = e^{j2\pi (p-1) \psi_{\rmd, \nT}^{\rmt}}$, $a_{\rmr, \nT, l}^{\rmt} = e^{-j2\pi (l - 1) \psi_{\rmr, \nT}^{\rmt}}$, and $\boldA_{\rms}(\psi_{\rms, \nT}^{\rmt})$ is the spatial target response matrix. Moreover, the spatial target response can be expressed as $\boldA_{\rms}(\psi_{\rms, \nT}^{\rmt}) = \bolda_{\rms, \rmr}(\psi_{\rms, \nT}^{\rmt}) \bolda_{\rms, \rmt}^{\mathrm{T}}(\psi_{\rms, \nT}^{\rmt})$, where $\bolda_{\rms, \rmr}(\psi_{\rms}) = [1, \dots, e^{j2\pi (\MR - 1)\psi_{\rms}}]^{\mathrm{T}}$ and $\bolda_{\rms, \rmt}(\psi_{\rms}) = [1, \dots, e^{j2\pi (\MT - 1)\psi_{\rms}}]^{\mathrm{T}}$ are the transmitting and receiving spatial steering vectors, respectively. 

Moreover, we consider $\NS$ stationary scatterers in the environment. The angle and range of the $\nS$th scatterer are denoted as $\theta_{\nS}^{\rms}$ and $r_{\nS}^{\rms}$, respectively. The corresponding range and spatial frequency is given by $\psi_{\rmr, \nS}^{\rms} = 2 r_{\nS}^{\rms} \Delta f / c$ and $\psi_{\rms, \nS}^{\rms} = d \sin\theta_{\nS}^{\rms} / \lambda$. The sensing channel of the $\nS$th scatter on the $l$th subcarrier is then
\begin{align}
\boldH_{\nS, l}^{\rms} = \alpha_{\nS}^{\rms} a_{\rmr, \nS, l}^{\rms} \boldA_{\rms}(\psi_{\rms, \nT}^{\rms}),
\end{align}
where $\alpha_{\nS}^{\rms}$ is the $\nS$th scatter's reflection coefficient, and $a_{\rmr, \nS, l}^{\rms} = e^{-j2\pi(l-1)\psi_{\rmr, \nS}^{\rms}}$. 

Hence, the overall sensing channel matrix can be written as
\begin{align}
  \boldH_{l, p} = \sum_{\nT \in \calN_{\rmt}} \boldH_{\nT, l, p}^{\rmt} + \sum_{\nS \in \calN_{\rms}} \boldH_{\nS, l}^{\rms},
\end{align}
where $\calN_{\rmt} = \{ 1, \dots, \NT \}$ and $\calN_{\rms} = \{ 1, \dots, \NS \}$. The corresponding receiving signal of the $\beamidx$th scan on the $l$th subcarrier at the $p$th symbol can be expressed as 
\begin{align}
  \label{eq:y-s-l-p-v1}
  \boldy_{\beamidx, l, p} &= \boldH_{l, p} \boldx_{\beamidx, l, p} + \boldn_{\beamidx, l, p} \notag \\
  &= \sum_{\nT \in \calN_{\rmt}} \alpha_{\nT}^{\rmt} a_{\rmd, \nT, p}^{\rmt} a_{\rmr, \nT, l}^{\rmt} g_{\beamidx, \nT, l, p}^{\rmt} \bolda_{\rms, \rmr}(\psi_{\rms, \nT}^{\rmt}) \notag \\
  &\hphantom{=} + \sum_{\nS \in \calN_{\rms}}\alpha_{\nS}^{\rms} a_{\rmr, \nS, l}^{\rms} g_{\beamidx, \nT, l, p}^{\rms} \bolda_{\rms, \rmr}(\psi_{\rms, \nS}^{\rms}) + \boldn_{\beamidx, l, p},
\end{align}
where $\boldx_{\beamidx, l, p}$ is the transmitting signal, $g_{\beamidx, \nT, l, p}^{\rmt} = \bolda_{\rms, \rmt}^{\mathrm{T}}(\psi_{\rms, \nT}^{\rmt})\boldx_{\beamidx, l, p}$, $g_{\beamidx, \nS, l, p}^{\rms} = \bolda_{\rms, \rmt}^{\mathrm{T}}(\psi_{\rms, \nS}^{\rms})\boldx_{\beamidx, l, p}$, and $\boldn_{\beamidx, l, p} \sim \calC\calN(\boldzero, \sigma^{2}\boldI)$.

\section{Moving Targets Sensing}

The procedure of the proposed sensing framework is characterized as following steps
\begin{enumerate}
    \item Steer the scanning beam towards $\tilde{\theta}_{1}, \dots, \tilde{\theta}_{\beamIdx}$, and transmit probing signals $\{ \boldx_{1, l, p} \}_{l, p}, \dots, \{ \boldx_{\beamIdx, l, p} \}_{l, p}$.
    \item Receive the echoed signal  $\{ \boldy_{1, l, p} \}_{l, p}, \dots, \{ \boldy_{\beamIdx, l, p} \}_{l, p}$, and remove the clutter from the receiving signal of each scan.
    \item Search for moving target candidates.
    \item Estimate the parameters of moving target candidates.
    \item Reject false candidates via target detection.
\end{enumerate}
In this section, we discuss the searching, parameter estimation and detection of moving target candidates.

\label{sec:moving-target-sensing}
\subsection{Clutter Removal and Target Searching}

To simplify the expression of the echo signal, we define $\beamset_{\nT} = \{ \beamidx \vert \theta_{\nT}^{\rmt} \in \Theta_{\beamidx} \}$ as the set of indices of coverage regions that covers the $\nT$th target, and $\calN_{\rms, \beamidx} = \{ \nS \vert \theta_{\nS}^{\rms} \in \Theta_{\beamidx} \}$ as the set of indices of scatters that lie in the coverage region of the $\beamidx$th scanning beam.

For any $\nT \in \calN_{\rmt}$, we focus on the echo signals with indices $\beamidx \in \calB_{\nT}$. We assume that the scanning beam is narrow enough, and thus the approximations $\tilde{\psi}_{\rms, \beamidx} \approx \psi_{\rms, \nT}^{\rmt} \approx \psi_{\rms, \nS}^{\rms}, \forall \nS \in \calN_{\rms, \beamidx}, \forall \beamidx \in \beamset_{\nT}$ hold. With these approximations, we have
\begin{align}
    g_{\beamidx, \nT^{\prime}, l, p}^{\rmt} &= \bolda_{\rms, \rmt}^{\mathrm{T}}(\psi_{\rms, \nT^{\prime}}^{\rmt}) \boldx_{\beamidx, l, p} \notag \\
    &\approx \underset{\tilde{g}_{\beamidx, l, p}}{ \underbrace{ \bolda_{\rms, \rmt}^{\mathrm{T}}(\tilde{\psi}_{\rms, \beamidx}) \boldx_{\beamidx, l, p} } } \delta(\nT - \nT^{\prime}), \ \beamidx \in \beamset_{\nT}, \label{eq:g-b-nT-prime-l-p-t} \\
    g_{\beamidx, \nS, l, p}^{\rms} &= \bolda_{\rms, \rmt}^{\mathrm{T}}(\psi_{\rms, \nS}^{\rms}) \boldx_{\beamidx, l, p} \notag \\
    &\approx \tilde{g}_{\beamidx, l, p} \sum_{\nS^{\prime} \in \calN_{\rms, s}} \delta(\nS - \nS^{\prime}), \beamidx \in \beamset_{\nT},
\end{align}
where $\tilde{\psi}_{\rms, \beamidx} = d \sin\tilde{\theta}_{\beamidx} / \lambda$, and $\tilde{g}_{\beamidx, l, p}$ is defined as the corresponding item. The receiving signal model~\cref{eq:y-s-l-p-v1} can be rewritten as
\begin{align}
    \label{eq:receiving-signal-single-target}
    \boldy_{\beamidx, l, p} &\approx \alpha_{\nT}^{\rmt} a_{\rmd, \nT, p}^{\rmt} a_{\rmr, \nT, l}^{\rmt} \tilde{g}_{\beamidx, l, p} \bolda_{\rms, \rmr}(\psi_{\rms, \nT}^{\rmt})\\
    &\hphantom{\approx} + \sum_{\nS \in \calN_{\rms, s}} \alpha_{\nS}^{\rms} a_{\rmr, \nS, l}^{\rms} \tilde{g}_{\beamidx, l, p} \bolda_{\rms, \rmr}(\psi_{\rms, \nS}^{\rms}) + \boldn_{\beamidx, l, p}. \notag
\end{align}
Since the scanning angle $\tilde{\theta}_{\beamidx}$ is known, it is easy to compute $\tilde{g}_{\beamidx, l, p}$ with~\cref{eq:g-b-nT-prime-l-p-t}. Defining $\tilde{\boldy}_{\beamidx, l, p} = \boldy_{\beamidx, l, p} / \tilde{g}_{\beamidx, l, p}$ and using~\cref{eq:receiving-signal-single-target}, we have
\begin{align}
    \label{eq:receiving-signal-single-target-bf-symbol-removal}
    \tilde{\boldy}_{\beamidx, l, p} &\approx \alpha_{\nT}^{\rmt} a_{\rmd, \nT, p}^{\rmt} a_{\rmr, \nT, l}^{\rmt} \bolda_{\rms, \rmr}(\psi_{\rms, \nT}^{\rmt}) \notag \\
    &\hphantom{\approx} + \sum_{\nS \in \calN_{\rms, s}} \alpha_{\nS}^{\rms} a_{\rmr, \nS, l}^{\rms} \bolda_{\rms, \rmr}(\psi_{\rms, \nS}^{\rms}) + \tilde{\boldn}_{\beamidx, l, p},
\end{align}
where $\tilde{\boldn}_{\beamidx, l, p} = 1 / \tilde{g}_{\beamidx, l, p} \boldn_{\beamidx, l, p}$. The first term represents target reflection, and the second term is caused by unknown scatters. 

It is seen from~\cref{eq:receiving-signal-single-target-bf-symbol-removal} that the clutter is invariant along the $p$-axis, while the echo is proportional to a sine function with frequency $\psi_{\rmd, \nT}^{\rmt}$. Hence, we propose to remove the clutter by performing high-pass filtering along the $p$-axis. The transfer function $H(z)$ of the infinite impulse response (IIR) filter can be directly derived from a well-designed analog high-pass filter, e.g., a Butterworth high-pass filter, via the bi-linear transform~\cite{Oppenheim2020Discrete}. After high-pass filtering, we get the output as follows:
\begin{align}
    \label{eq:check-y-s-nt-star-l-p}
    \check{\boldy}_{\beamidx, l, p} \approx \alpha_{\nT}^{\rmt} \check{\alpha}_{\nT} a_{\rmd, \nT, p}^{\rmt} a_{\rmr, \nT, l}^{\rmt} \bolda_{\rms, \rmr}(\psi_{\rms, \nT}^{\rmt}) + \check{\boldn}_{\beamidx, l, p},
\end{align}
where $\check{\alpha}_{\nT}$ with $\lvert \check{\alpha}_{\nT} \rvert \approx 1$ is the coefficient of the high-pass filter. Hence, the scanning directions towards potential moving targets can be identified by finding the peaks of the following spectrum
\begin{align}
    P({\beamidx}) = \frac{1}{LP}\sum_{l=1}^{L}\sum_{p=1}^{P} \lVert \check{\boldy}_{\beamidx, l, p} \rVert^{2}.
\end{align}

\subsection{Target Parameter Estimation}

Let $\hat{\beamset}$ be the set of indices of peaks of $P_{\beamidx}$. We need to estimate the kinematic parameter of each moving target candidate from receiving signal $\check{\boldy}_{\hat{\beamidx}, l, p}$ with index $\hat{\beamidx} \in \hat{\beamset}$.

We first briefly introduce the problem of estimating the frequency of a sinusoidal signal. Then, we show that the kinematic parameter estimation problem can be recast as the frequency estimation problem. 

Consider the following $M$-dimensional noisy observations
\begin{align}
    \label{eq:observation-model}
    \boldF = \begin{bmatrix}
        \alpha_{1} \boldf & \dots & \alpha_{I}\boldf
    \end{bmatrix} + \boldN \in \bbC^{M \times I},
\end{align}
where $\alpha_{i}$ is an unknown coefficient, $\boldf = [f_{1}, \dots, f_{M}]^{\mathrm{T}}$ with $f_{m} = e^{j2 \pi (m-1) \psi}$ being a sinusoidal signal with frequency $\psi$, $\boldN$ is the zero-mean Gaussian noise. The estimating procedure of the well-known root-MUSIC algorithm is summarized in \Cref{alg:root-music}.
\begin{algorithm}[H]
    \caption{Root-MUSIC for estimating frequency}
    \label{alg:root-music}

    \SetKwRepeat{Do}{do}{while}
    \setcounter{AlgoLine}{0}
    \SetKw{Continue}{continue}
    \SetKw{Break}{break}
    \SetKwInOut{Input}{Input}
    \SetKwInOut{Output}{Output}
    \SetKwComment{tcp}{\% }{}

    \Input{ Signal matrix $\boldF \in \bbC^{P \times I}$. }
    
    Compute the $I-1$-dimensional noise subspace $\boldV_{\rmn}$ using the sample covariance matrix $\hat{\boldR} = 1 / I \boldF \boldF^{\mathrm{H}}$.

    Construct the polynomial function $f(z) = \boldp^{\mathrm{H}}(z) \boldV_{\rmn}\boldV_{\rmn}^{\mathrm{H}} \boldp(z)$, and find the root $z^{\star}$ of $f(z)$ which is inside and closest to the unit circle.

    \Output{ Estimated frequency $\hat{\psi} = \arg\left( z^{\star} \right)$. }
\end{algorithm}

\subsubsection{Angle Estimation} Using~\cref{eq:check-y-s-nt-star-l-p}, we have
\begin{align}
    \label{eq:check-y-s-hat-star-l-p}
    \check{\boldy}_{\hat{\beamidx}, l, p} \approx \alpha_{\nT}^{\rmt} \check{\alpha}_{\nT} a_{\rmd, \nT, p}^{\rmt} a_{\rmr, \nT, l}^{\rmt} \bolda_{\rms, \rmr}(\psi_{\rms, \nT}^{\rmt}) + \check{\boldn}_{\hat{\beamidx}, l, p}.
\end{align}
Note that for different $l$ and $p$, the post-processing signals $\check{\boldy}_{\hat{\beamidx}, l, p}$ have the same sinusoidal component $\bolda_{\rms, \rmr}(\psi_{\rms, \nT}^{\rmt})$. We then form a matrix $\check{\boldY}_{\rms, \hat{\beamidx}}$ as follows
\begin{align}
    \label{eq:check-boldY-s-shat}
    \check{\boldY}_{\rms, \hat{\beamidx}} = [\check{\boldy}_{\hat{\beamidx}, 1, 1}, \dots, \check{\boldy}_{\hat{\beamidx}, L, 1}, \dots, \check{\boldy}_{\hat{\beamidx}, 1, P}, \dots, \check{\boldy}_{\hat{\beamidx}, L, P}].
\end{align}
Then, the spatial frequency $\hat{\psi}_{\rms, \nT}^{\rmt}$ can be estimated by invoking \Cref{alg:root-music}, and we can compute the direction of the target using $\hat{\theta}_{\nT}^{\rmt} = \arcsin\left( \lambda \hat{\psi}_{\rms, \nT}^{\rmt} / d \right)$.

\subsubsection{Range Estimation}

To facilitate range estimation, we first rearrange the elements as
\begin{align}
    \check{\boldy}_{\rmr, \hat{\beamidx}, \mR, p} &= [[\check{\boldy}_{\hat{\beamidx}, 1, p}]_{\mR}, \dots, [\check{\boldy}_{\hat{\beamidx}, L, p}]_{\mR}]^{\mathrm{T}} \\
    &= \alpha_{\nT}^{\rmt} \check{\alpha}_{\nT}a_{\rmd, \nT, p}^{\rmt} a_{\rms, \rmr, \mT}^{\rmt} \bolda_{\rmr}(\psi_{\rmr, \nT}^{\rmt}) + \check{\boldn}_{\rmr, \hat{\beamidx}, \mR, p}, \notag
\end{align}
where $[\check{\boldy}_{\hat{\beamidx}, l, p}]_{\mR}$ is the $\mR$th entry of $\check{\boldy}_{\hat{\beamidx}, l, p}$, and $\bolda_{\rmr}(\psi_{\rmr}) = [1, \dots, e^{-j2\pi(L-1)\psi_{\rmr}}]^{\mathrm{T}}$ stands for the range steering vector. Then, we form the signal matrix as follows
\begin{align}
    \check{\boldY}_{\rmr, \hat{\beamidx}} = [\check{\boldy}_{\rmr, \hat{\beamidx}, 1, 1}, \dots, \check{\boldy}_{\rmr, \hat{\beamidx}, \MR, 1}, \check{\boldy}_{\rmr, \hat{\beamidx}, 1, P}, \dots, \check{\boldy}_{\rmr, \hat{\beamidx}, \MR, P}].
\end{align}
It can be seen that $\check{\boldY}_{\rmr, \hat{\beamidx}}$ has the same form as~\cref{eq:observation-model}. Thus, we can estimate the range frequency $\hat{\psi}_{\rmr, \nT}^{\rmt}$ by invoking \Cref{alg:root-music} again. Then, the range estimate is given by $\hat{r}_{\nT}^{\rmt} = c \hat{\psi}_{\rmr, \nT}^{\rmt} / (2 \Delta f)$.

\subsubsection{Doppler Estimation}

Similarly, to estimate the Doppler frequency, we rearrange the elements and obtain
\begin{align}
    \check{\boldy}_{\rmd, \hat{\beamidx}, \mR, l} &= [[\check{\boldy}_{\hat{\beamidx}, l, 1}]_{\mR}, \dots, [\check{\boldy}_{\hat{\beamidx}, l, P}]_{\mR}]^{\mathrm{T}} \\
    &= \alpha_{\nT}^{\rmt} \check{\alpha}_{\nT}a_{\rmr, \nT, l}^{\rmt} a_{\rms, \rmr, \mT}^{\rmt} \bolda_{\rmd}(\psi_{\rmd, \nT}^{\rmt}) + \check{\boldn}_{\rmd, \hat{\beamidx}, \mR, l}, \notag
\end{align}
where $\bolda_{\rmd}(\psi_{\rmd}) = [1, \dots, e^{j2\pi(P-1)\psi_{\rmd}}]^{\mathrm{T}}$ represents the Doppler steering vector. Then, we form the signal matrix as follows
\begin{align}
    \check{\boldY}_{\rmd, \hat{\beamidx}} = [\check{\boldy}_{\rmd, \hat{\beamidx}, 1, 1}, \dots, \check{\boldy}_{\rmd, \hat{\beamidx}, \MR, 1}, \check{\boldy}_{\rmd, \hat{\beamidx}, 1, P}, \dots, \check{\boldy}_{\rmd, \hat{\beamidx}, \MR, L}].
\end{align}
We can observe that $\check{\boldY}_{\rmd, \hat{\beamidx}}$ has the same form as~\cref{eq:observation-model}, and thus we can estimate the radial velocity using $\hat{v}_{\nT}^{\rmt} = \lambda \hat{\psi}_{\rmr, \nT}^{\rmt} / (2 T)$, where the Doppler frequency is estimated by invoking \Cref{alg:root-music}.

\subsubsection{Cramér-Rao Bound}
Denote $\boldeta^{\rmt} = [\theta_{1}^{\rmt}, \dots \theta_{\NT}^{\rmt},r_{1}^{\rmt}, \dots r_{\NT}^{\rmt},v_{1}^{\rmt}, \dots v_{\NT}^{\rmt}]^{\mathrm{T}}$ and $\boldeta^{\rms} = [\theta_{1}^{\rms}, \dots \theta_{\NS}^{\rms},r_{1}^{\rms}, \dots r_{\NS}^{\rms}]^{\mathrm{T}}$ as the kinematic parameter vectors of moving targets and scatters, $\boldalpha^{\rmt} = [\alpha_{1}^{\rmt}, \dots, \alpha_{\NT}^{\rmt}]^{\mathrm{T}}$, $\boldalpha^{\rms} = [\alpha_{1}^{\rms}, \dots, \alpha_{\NS}^{\rms}]^{\mathrm{T}}$, and $\boldalpha = [(\boldalpha^{\rmt})^{\mathrm{T}}, (\boldalpha^{\rms})^{\mathrm{T}}]^{\mathrm{T}}$ as the reflection coefficient vectors, respectively.

Defining $\bolda_{\beamidx, \nT, l, p}^{\rmt} = a_{\rmd, \nT, p}^{\rmt} a_{\rmr, \nT, l}^{\rmt} g_{\beamidx, \nT, l, p}^{\rmt} \bolda_{\rms, \rmr}(\psi_{\rms, \nT}^{\rmt})$ and $\bolda_{\beamidx, \nS, l, p}^{\rms} = a_{\rmd, \nS, p}^{\rms} a_{\rmr, \nS, l}^{\rms} g_{\beamidx, \nS, l, p}^{\rms} \bolda_{\rms, \rmr}(\psi_{\rms, \nS}^{\rms})$, we can rewrite~\cref{eq:y-s-l-p-v1} as
\begin{align}
    \boldy_{\beamidx, l, p} = \boldA_{\beamidx, l, p}^{\rmt}\boldalpha^{\rmt} + \boldA_{\beamidx, l, p}^{\rms}\boldalpha^{\rms} + \boldn_{\beamidx, l, p},
\end{align}
where
\begin{align}
    \boldA_{\beamidx, l, p}^{\rmt} = \begin{bmatrix}
        \bolda_{\beamidx, 1, l, p}^{\rmt} & \dots & \bolda_{\beamidx, \NT, l, p}^{\rmt}
    \end{bmatrix}, \\
    \boldA_{\beamidx, l, p}^{\rms} = \begin{bmatrix}
        \bolda_{\beamidx, 1, l, p}^{\rms} & \dots & \bolda_{\beamidx, \NS, l, p}^{\rms}
    \end{bmatrix}.
\end{align}

Using the echo of the $\beamidx$th scan for estimation, the log-likelihood can be expressed as  
\begin{align}
    \calL_{\beamidx} = -\MR L P \log \sigma^{2} - \frac{1}{\sigma^{2}} \sum_{l, p}\lVert \bolde_{\beamidx, l, p} \rVert^{2},
\end{align}
where $\bolde_{\beamidx, l, p} = \boldy_{\beamidx, l, p} - \boldn_{\beamidx, l, p}$.

The first-order derivatives with respect to unknown parameters of the log-likelihood can be calculated as
\begin{align}
    \frac{\partial \calL_{\beamidx}}{\partial \sigma^{2}} &= -\frac{\MR P L}{\sigma^{2}} + \frac{1}{(\sigma^{2})^{2}} \sum_{l, p} \lVert \bolde_{\beamidx, l, p} \rVert^{2}, \\
    \frac{\partial \calL_{\beamidx}}{\partial \boldeta^{\rmt}} &= \frac{2}{\sigma^{2}} \sum_{l, p} \Re{ (\dot{\boldA}_{\beamidx, l, p}^{\rmt} \boldD_{\boldalpha}^{\rmt})^{\mathrm{H}} \bolde_{\beamidx, l, p} }, \\
    \frac{\partial \calL_{\beamidx}}{\partial \boldeta^{\rms}} &= \frac{2}{\sigma^{2}} \sum_{l, p} \Re{ (\dot{\boldA}_{\beamidx, l, p}^{\rms} \boldD_{\boldalpha}^{\rms})^{\mathrm{H}} \bolde_{\beamidx, l, p} },
\end{align}
\begin{align}
    \frac{\partial \calL_{\beamidx}}{\partial \Re{\boldalpha^{\rmt}}} &= \sum_{l, p} \Re{ (\boldA_{\beamidx, l, p}^{\rmt})^{\mathrm{H}} \bolde_{\beamidx, l, p} }, \\
    \frac{\partial \calL_{\beamidx}}{\partial \Im{\boldalpha^{\rmt}}} &= \sum_{l, p} \Im{ (\boldA_{\beamidx, l, p}^{\rmt})^{\mathrm{H}} \bolde_{\beamidx, l, p} }, \\
    \frac{\partial \calL_{\beamidx}}{\partial \Re{\boldalpha^{\rms}}} &= \sum_{l, p} \Re{ (\boldA_{\beamidx, l, p}^{\rms})^{\mathrm{H}} \bolde_{\beamidx, l, p} }, \\
    \frac{\partial \calL_{\beamidx}}{\partial \Im{\boldalpha^{\rms}}} &= \sum_{l, p} \Im{ (\boldA_{\beamidx, l, p}^{\rms})^{\mathrm{H}} \bolde_{\beamidx, l, p} },
\end{align}
where
\begin{align*}
    \dot{\boldA}_{\beamidx, l, p}^{\rmt} &= \left[
        \frac{\partial \bolda_{\beamidx, 1, l, p}^{\rmt}}{\partial \theta_{1}^{\rmt}},\dots,\frac{\partial \bolda_{\beamidx, 1, l, p}^{\rmt}}{\partial \theta_{\NT}^{\rmt}},\frac{\partial \bolda_{\beamidx, 1, l, p}^{\rmt}}{\partial r_{1}^{\rmt}},\dots,\frac{\partial \bolda_{\beamidx, 1, l, p}^{\rmt}}{\partial r_{\NT}^{\rmt}},\right. \notag \\
        &\hphantom{=[}\left.\frac{\partial \bolda_{\beamidx, 1, l, p}^{\rmt}}{\partial v_{1}^{\rmt}},\dots,\frac{\partial \bolda_{\beamidx, 1, l, p}^{\rmt}}{\partial v_{\NT}^{\rmt}}\right], \\
        \boldD_{\boldalpha}^{\rmt} &= [\diag{\boldalpha^{\rmt}},\diag{\boldalpha^{\rmt}},\diag{\boldalpha^{\rmt}}], \\
        \dot{\boldA}_{\beamidx, l, p}^{\rms} &= \left[
        \frac{\partial \bolda_{\beamidx, 1, l, p}^{\rms}}{\partial \theta_{1}^{\rms}},\dots,\frac{\partial \bolda_{\beamidx, 1, l, p}^{\rms}}{\partial \theta_{\NS}^{\rms}}, \frac{\partial \bolda_{\beamidx, 1, l, p}^{\rms}}{\partial r_{1}^{\rms}},\dots,\frac{\partial \bolda_{\beamidx, 1, l, p}^{\rms}}{\partial r_{\NS}^{\rms}}\right], \\
        \boldD_{\boldalpha}^{\rms} &= [\diag{\boldalpha^{\rms}},\diag{\boldalpha^{\rms}}].
\end{align*}


For notional convenience, we use $\boldF_{\beamidx, \boldx, \boldy}$ to denote the partial Fisher information matrix (FIM) $\bbE\left[ \frac{\partial \calL_{\beamidx}}{\partial \boldx}\left( \frac{\partial \calL_{\beamidx}}{\partial \boldy} \right)^{\mathrm{T}} \right]$. They are
\begin{align}
    &\boldF_{\beamidx, \sigma^{2}, \sigma^{2}} = \frac{\MR L P}{(\sigma^{2})^{2}}, \\
    &\boldF_{\beamidx, \sigma^{2}, (\cdot)} = \boldzero, \\
    &\boldF_{\beamidx, \boldeta^{x}, \boldeta^{y}} = \frac{2}{\sigma^{2}} \sum_{l, p} \Re{ (\dot{\boldA}_{\beamidx, l, p}^{x} \boldD_{\boldalpha}^{x})^{\mathrm{H}} \dot{\boldA}_{\beamidx, l, p}^{y} \boldD_{\boldalpha}^{y} }, \\
    &\boldF_{\beamidx, \boldeta^{x}, \Re{ \boldalpha }} = \frac{2}{\sigma^{2}}\sum_{l, p} \Re{ (\boldA_{\beamidx, l, p}^{x} \boldD_{\boldalpha}^{x})^{\mathrm{H}} \boldA_{\beamidx, l, p} }, \\
    &\boldF_{\beamidx, \boldeta^{x}, \Im{ \boldalpha }} = -\frac{2}{\sigma^{2}}\sum_{l, p} \Im{ (\boldA_{\beamidx, l, p}^{x} \boldD_{\boldalpha}^{x})^{\mathrm{H}} \boldA_{\beamidx, l, p} }, \\
    &\boldF_{\beamidx, \Re{\boldalpha}, \Re{\boldalpha}} = \frac{2}{\sigma^{2}}\sum_{l, p} \Re{ \boldA_{\beamidx, l, p}^{\mathrm{H}} \boldA_{\beamidx, l, p} }, \\
    &\boldF_{\beamidx, \Im{\boldalpha}, \Im{\boldalpha}} = \frac{2}{\sigma^{2}}\sum_{l, p} \Re{ \boldA_{\beamidx, l, p}^{\mathrm{H}} \boldA_{\beamidx, l, p} }, \\
    &\boldF_{\beamidx, \Re{\boldalpha}, \Im{\boldalpha}} = -\frac{2}{\sigma^{2}}\sum_{l, p} \Im{ \boldA_{\beamidx, l, p}^{\mathrm{H}} \boldA_{\beamidx, l, p} }, 
\end{align}
where $\boldA_{\beamidx, l, p} = [\boldA_{\beamidx, l, p}^{\rmt}, \boldA_{\beamidx, l, p}^{\rms}]$, and $x, y \in \{ \rmt, \rms \}$.

Since the partial FIMs $\boldF_{\beamidx, \sigma^{2}, (\cdot)}$ are zero, the matrix inversion lemma gives 
\begin{align}
    \label{eq:CRB-v1}
    \CRB_{\beamidx, \boldeta^{\rmt}} = \boldF_{\beamidx, \boldeta^{\rmt}, \boldeta^{\rmt}}^{-1} = \boldF_{\beamidx, 1} - \boldF_{\beamidx, 2}\boldF_{\beamidx, 3}^{-1}\boldF_{\beamidx, 2},
\end{align}
where
\begin{align}
    \boldF_{\beamidx, 1} &= \frac{2}{\sigma^{2}} \sum_{l, p}\Re{(\boldJ_{\beamidx, l, p}^{1})^{\mathrm{H}} \boldJ_{\beamidx, l, p}^{1}}, \\
    \boldF_{\beamidx, 2} &= \frac{2}{\sigma^{2}} \sum_{l, p} [\Re{\boldJ_{\beamidx, l, p}^{1}}, -\Im{\boldJ_{\beamidx, l, p}^{1}}], \\
    \boldF_{\beamidx, 3} &= \sum_{l, p}\frac{2}{\sigma^{2}} \begin{bmatrix}
        \Re{\boldJ_{\beamidx, l, p}^{2}} & -\Im{\boldJ_{\beamidx, l, p}^{2}} \\
        \Im{\boldJ_{\beamidx, l, p}^{2}} & \Re{\boldJ_{\beamidx, l, p}^{2}}
    \end{bmatrix}
\end{align}
with
\begin{align}
    \boldJ_{\beamidx, l, p}^{1} &= [\dot{\boldA}_{\beamidx, l, p}^{\rmt} \boldD_{\boldalpha}^{\rmt},\dot{\boldA}_{\beamidx, l, p}^{\rms} \boldD_{\boldalpha}^{\rms}], \\
    \boldJ_{\beamidx, l, p}^{2} &= \boldA_{\beamidx, l, p}^{\mathrm{H}} \boldA_{\beamidx, l, p}.
\end{align}

\subsection{Target Detection}
\label{sec:target-detection}
The moving target candidates obtained in the beam scanning stage may contain false targets, and we employ the target detection technique to reject false alarms. Specifically, we consider an arbitrary target candidate, and assume that the corresponding beam index is $\beamidx$. We assume that the estimated parameters are $(v, r, \theta)$, and denote $(\psi_{\rmd}, \psi_{\rmr}, \psi_{\rms})$ as the corresponding Doppler, range and spatial frequency. To facilitate detection, we sample a range-spatial frequency grid $\calG_{\beamidx} = \{ (\tilde{\psi}_{\rmr, \nG}, \tilde{\psi}_{\rms, \beamidx, \nG}) \}_{\nG = 1}^{\NG}$ within the coverage region of the $\beamidx$th beam. Assuming that the clutter falls within a subspace spanned by the columns of $\tilde{\boldA}_{\beamidx, l, p} = [\tilde{\bolda}_{\beamidx, l, p, 1}, \dots, \tilde{\bolda}_{\beamidx, l, p, \NG}]$, where $\tilde{\bolda}_{\beamidx, l, p, \nG} = \tilde{g}_{\beamidx, l, p}e^{-j 2 \pi (l - 1) \tilde{\psi}_{\rmr, \nG}} \bolda_{\rms, \rmr}(\tilde{\psi}_{\rms, \beamidx, \nG})$, the target detection problem can be formulated as
\begin{align}
    \begin{cases}
        \boldy_{\beamidx, l, p} = \tilde{\boldA}_{\beamidx, l, p} \tilde{\boldalpha}_{\beamidx} + \boldn_{\beamidx, l, p}, & \calH_{0}, \\
        \boldy_{\beamidx, l, p} = \alpha \bolda_{\beamidx, l, p} + \tilde{\boldA}_{\beamidx, l, p} \tilde{\boldalpha}_{\beamidx} + \boldn_{\beamidx, l, p}, & \calH_{1},
    \end{cases}
\end{align}
where $\calH_{0}$ is the null hypothesis, $\calH_{1}$ is the alternative hypothesis, $\alpha$ is the unknown magnitude, and $\bolda_{\beamidx, l, p} = \tilde{g}_{\beamidx, l, p}e^{j2\pi (p-1) \psi_{\rmd}} e^{-j2\pi(l-1)\psi_{\rmr}}\bolda_{\rms, \rmr}(\psi_{\rms})$. The likelihoods under these hypotheses are given by
\begin{align}
    p_{0} &= \varrho \exp\left( -\frac{1}{\sigma^{2}} \sum_{l, p}\lVert \boldy_{\beamidx, l, p} - \tilde{\boldA}_{\beamidx, l, p} \tilde{\boldalpha}_{\beamidx} \rVert^{2} \right), \\
    p_{1} &= \varrho \exp\left( -\frac{1}{\sigma^{2}} \sum_{l, p}\lVert \boldy_{\beamidx, l, p} - \alpha\bolda_{\beamidx, l, p} - \tilde{\boldA}_{\beamidx, l, p} \tilde{\boldalpha}_{\beamidx} \rVert^{2} \right),
\end{align}
where $\varrho = (\pi \sigma^{2})^{-\MR L P}$ is the normalization factor.

The generalized likelihood ratio (GLR)~\cite{Kay1993Fundamentals} is then defined as
\begin{align}
    \label{eq:GLR-definition}
    t = \frac{ \max_{ \alpha, \tilde{\boldalpha}_{\beamidx}, \sigma^{2} } p_{1} }{ \max_{ \tilde{\boldalpha}_{\beamidx}, \sigma^{2} } p_{0} }.
\end{align}
To derive a closed-form GLR, we need to compute the maximum likelihood (ML) estimates of unknown parameters. The ML estimates under $\calH_{0}$ is given by
\begin{align}
    \hat{\tilde{\boldalpha}}_{\beamidx, 0} &= \sum_{l, p}\left( \tilde{\boldA}_{\beamidx, l, p}^{\mathrm{H}} \tilde{\boldA}_{\beamidx, l, p} \right)^{+} \tilde{\boldA}_{\beamidx, l, p}^{\mathrm{H}} \boldy_{\beamidx, l, p}, \label{eq:hat-alpha-0}\\
    \hat{\sigma}_{0}^{2} &= \frac{1}{\MR L P} \sum_{l, p}\boldy_{\beamidx, l, p}^{\mathrm{H}} \boldP_{\tilde{\boldA}_{\beamidx, l, p}}^{\perp} \boldy_{\beamidx, l, p}, \label{eq:sigma-2-0}
\end{align}
where $(\cdot)^{+}$ represents Moore–Penrose inverse~\cite{Zhang2020AMatrix}, $\boldP_{\tilde{\boldA}_{\beamidx, l, p}}^{\perp} = \boldI - \tilde{\boldA}_{\beamidx, l, p} \left( \tilde{\boldA}_{\beamidx, l, p}^{\mathrm{H}} \tilde{\boldA}_{\beamidx, l, p} \right)^{+} \tilde{\boldA}_{\beamidx, l, p}^{\mathrm{H}}$. The ML estimates uner $\calH_{1}$ can be expressed as
\begin{align}
    \hat{\tilde{\boldalpha}}_{\beamidx, 1} &= \sum_{l, p}\tilde{\boldA}_{\beamidx, l, p}^{+} \left( \boldy_{\beamidx, l, p} - \hat{\alpha}_{1} \bolda_{\beamidx, l, p} \right), \label{eq:hat-alpha-1}\\
    \hat{\alpha} &= \sum_{l, p}\frac{(\bolda_{\beamidx, l, p}^{\rmt})^{\mathrm{H}} \boldP_{\tilde{\boldA}_{\beamidx, l, p}}^{\perp} \boldy_{\beamidx, l, p}}{(\bolda_{\beamidx, l, p}^{\rmt})^{\mathrm{H}} \boldP_{\tilde{\boldA}_{\beamidx, l, p}}^{\perp} \bolda_{\beamidx, l, p}^{\rmt}}, \label{eq:hat-alpha}\\
    \hat{\sigma}_{1}^{2} &= \hat{\sigma}_{0}^{2} - \frac{1}{\MR L P}\sum_{l, p}\frac{\lvert \bolda_{\beamidx, l, p}^{\mathrm{H}} \boldP_{\tilde{\boldA}_{\beamidx, l, p}}^{\perp} \boldy_{\beamidx, l, p} \rvert^{2}}{\bolda_{\beamidx, l, p}^{\mathrm{H}} \boldP_{\tilde{\boldA}_{\beamidx, l, p}}^{\perp} \bolda_{\beamidx, l, p}}. \label{eq:sigma-2-1}
\end{align}

Inserting~\cref{eq:hat-alpha-0}-\cref{eq:sigma-2-1} into~\cref{eq:GLR-definition}, after some algebra, the detection criterion can be expressed as
\begin{align}
    \sum_{l, p}\frac{\lvert \bolda_{\beamidx, l, p}^{\mathrm{H}} \boldP_{\tilde{\boldA}_{\beamidx, l, p}}^{\perp} \boldy_{\beamidx, l, p} \rvert^{2}}{\MR L P\hat{\sigma}_{0}^{2} \bolda_{\beamidx, l, p}^{\mathrm{H}} \boldP_{\tilde{\boldA}_{\beamidx, l, p}}^{\perp} \bolda_{\beamidx, l, p}} \underset{\calH_{0}}{\overset{\calH_{1}}{\gtrless}} \gamma,
\end{align}
where $\gamma$ is the detection threshold.

\section{Simulation Results}
\label{sec:simulation-results}

In this section, we carry out simulations to evaluate the performance of the proposed methods. Unless otherwise specified, we set $\MR = 16$, $\MT = 64$, $L = 16$, $P = 20$, $f_{\rmc} = \qty{60}{\giga\hertz}$, $\Delta {f} = \qty{10}{\mega\hertz}$, $T_{\rmg} = \qty{0.2}{\milli\second}$, $d = \lambda / 2$. We assume that there are \num{2} moving targets and \num{400} scatters in the environment, whose parameters are distributed as $\theta_{\nT}^{\rmt}, \theta_{\nS}^{\rms} \sim \calU(\qty{-60}{\degree}, \qty{60}{\degree})$, $r_{\nT}^{\rmt}, r_{\nS}^{\rms} \sim \calU(\qty{1}{\meter}, \qty{7}{\meter})$, $v_{\nT}^{\rmt} \sim \calU(\qty{1}{\meter\per\second}, \qty{4}{\meter\per\second})$, $\alpha_{\nT}^{\rmt} \sim \calC\calN(0, 1)$, $\alpha_{\nS}^{\rms}\sim \calC\calN(0, 0.5)$, where $\calU$ represents the uniform distribution. All parameters are randomly generated and then fixed in simulations. The generated kinematic parameters of targets are listed in \Cref{tab:target-parameter}.
\begin{table}[t!]
    \centering
    \caption{Generated kinematic Parameters of Targets}
    \label{tab:target-parameter}
    \begin{tabular}{cccc}
    \toprule
    Target & Angle   & Range & Radial Velocity \\
    \midrule
    1      & \qty{-48.295}{\degree} & \qty{4.281}{\meter} & \qty{3.911}{\meter\per\second} \\
    2      & \qty{15.883}{\degree}  & \qty{2.670}{\meter} & \qty{1.473}{\meter\per\second} \\
    \bottomrule
    \end{tabular}
\end{table}

\begin{figure}[t!]
    \centering
    \includegraphics[width=.9\linewidth]{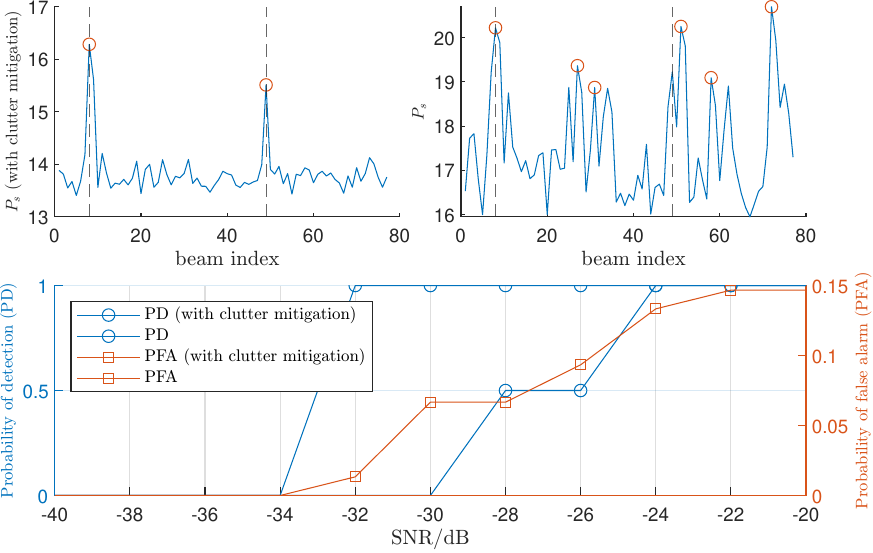}
    \caption{Comparison of moving target searching performance.}
    \label{fig:beam-selection}
\end{figure}

We first evaluate the target searching performance, and display the results in \Cref{fig:beam-selection}. In the visualization results, the signal-to-noise ratio (SNR) is set to \qty{-30}{\deci\bel}, and the dashed lines represent the scanning directions that are closest to the directions of moving targets. The results verify that the proposed clutter mitigation method is able to boost the sensing performance under strong clutter.

\begin{figure}[t!]
    \centering
    \includegraphics[width=.9\linewidth]{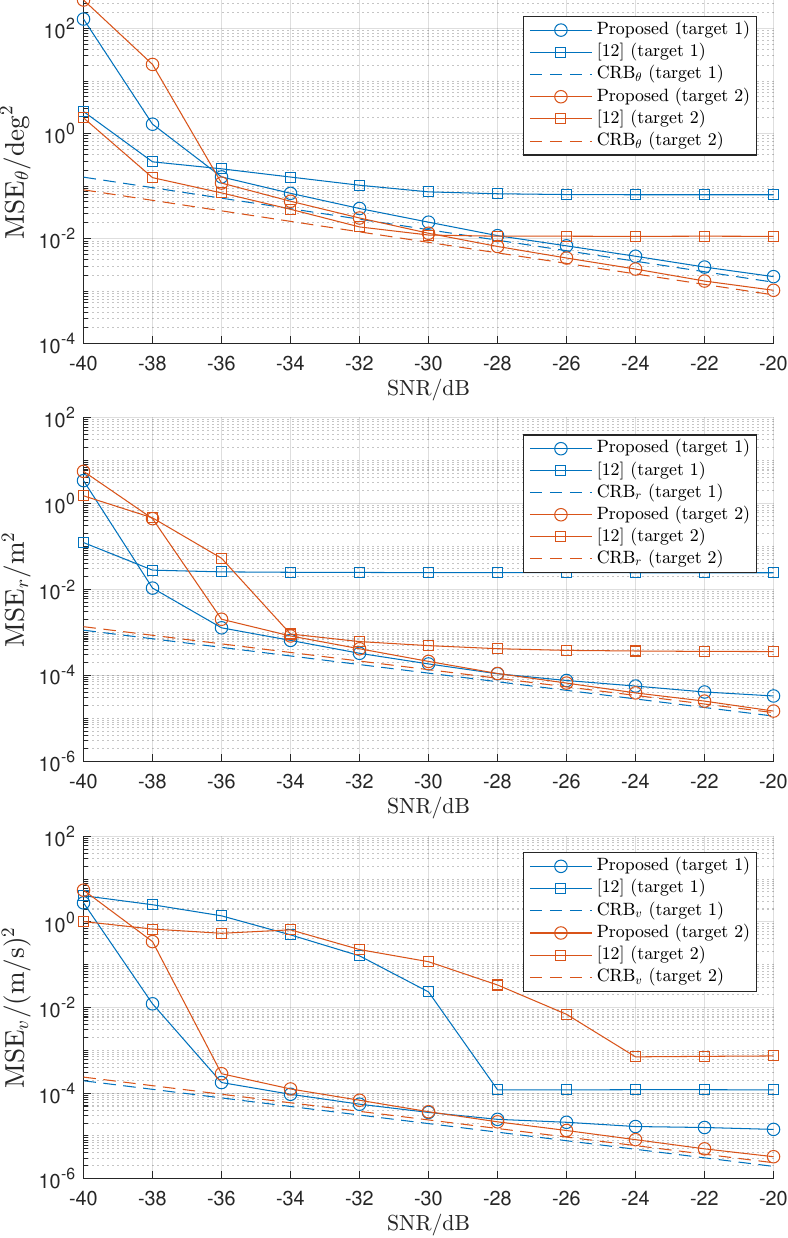}
    \caption{Comparison of the parameter estimation performance.}
    \label{fig:parameter-estimation-MSE}
\end{figure}
We then assume that the correct target candidates are detected, and focus on the parameter estimation performance. \Cref{fig:parameter-estimation-MSE} compares the proposed approach with the method in~\cite{Chen2023Multiple}. We can see that the proposed algorithm reaches the near-optimal performance asymptotically. Moreover, it is also observed that with the proposed clutter mitigation procedure, the sensing performance can be significantly improved especially in high-SNR regime, while the competitor's precision is limited by the strong clutter. 

\begin{figure}[t!]
    \centering
    \includegraphics[width=.85\linewidth]{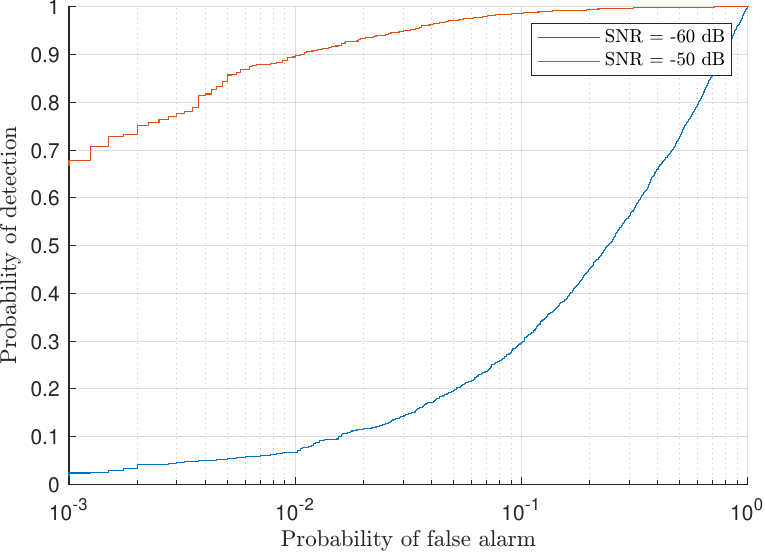}
    \caption{ROC curves with different SNR.}
    \label{fig:ROC}
\end{figure}

In \Cref{fig:ROC}, we report the receiver operating characteristic (ROC) curves to assess the detection performance for target 1. The results verify that the proposed algorithm is able to achieve a high detection probability with only a small probability of false alarm. Moreover, we observe that the performance degrades significantly in the low-SNR regime.

\section{Conclusion}
\label{sec:conclusion}

In this paper, we investigated the problem of moving target sensing in environments populated with stationary scatters, including clutter mitigation, target searching, target parameter estimation, and target detection. Specifically, we proposed the Doppler-domain high-pass filtering technique, the root-MUSIC-based approach, and the GLR-based detection algorithm. Simulation results have validated the effectiveness and high performance of the proposed methods.

\printbibliography

\end{document}